\documentclass[a4paper,11pt]{scrartcl}

\usepackage[T1]{fontenc}
\usepackage[utf8]{inputenc}

\usepackage[english]{babel}

\usepackage{lmodern}

\usepackage{fourier} 

\usepackage[a4paper,left=20mm,top=30mm,right=20mm,bottom=30mm]{geometry}
\usepackage[autostyle]{csquotes}
\usepackage[
	backend=biber,
	bibstyle=ieee,
	citestyle=numeric,
	natbib=true,
	date=year,
	giveninits=true,
	sortlocale=en_US,
	url=false, 
	doi=false,
	eprint=false,
	isbn=false
]{biblatex}
\usepackage{booktabs}
\usepackage{lipsum}
\usepackage{microtype}
\usepackage{url}
\usepackage{multicol}
\usepackage{nameref}
\usepackage{hyperref}
\hypersetup{pdfborder={0 0 0},
	colorlinks=true,
	linkcolor=black,
	citecolor=black,
	urlcolor=black}
\usepackage{graphicx}
\usepackage{float}
\usepackage{wrapfig}
\usepackage[labelfont=bf,position=bottom,format=plain]{caption}

\begin{document}

\begin{center}

{\Large \textbf{Wurtzite InP Microdisks: from Epitaxy to Room-Temperature Lasing}}

\vspace{0.1cm}

Philipp Staudinger\textsuperscript{+}, Svenja Mauthe\textsuperscript{+}, Noelia Vico Triviño, Steffen Reidt, Kirsten E. Moselund, and~Heinz~Schmid*

\vspace{0.1cm}

\emph{IBM Research Zurich, Säumerstrasse 4, 8803 Rüschlikon, Switzerland} \\
\emph{\textsuperscript{+} These authors contributed equally to this work.} \\
\emph{* Correspondence to sih@zurich.ibm.com}

\vspace{0.3cm}

\end{center}

\setlength{\columnsep}{15pt}
\begin{wrapfigure}{r}{0.35\textwidth}
	\vspace{-12pt}
	\includegraphics[width=0.35\textwidth]{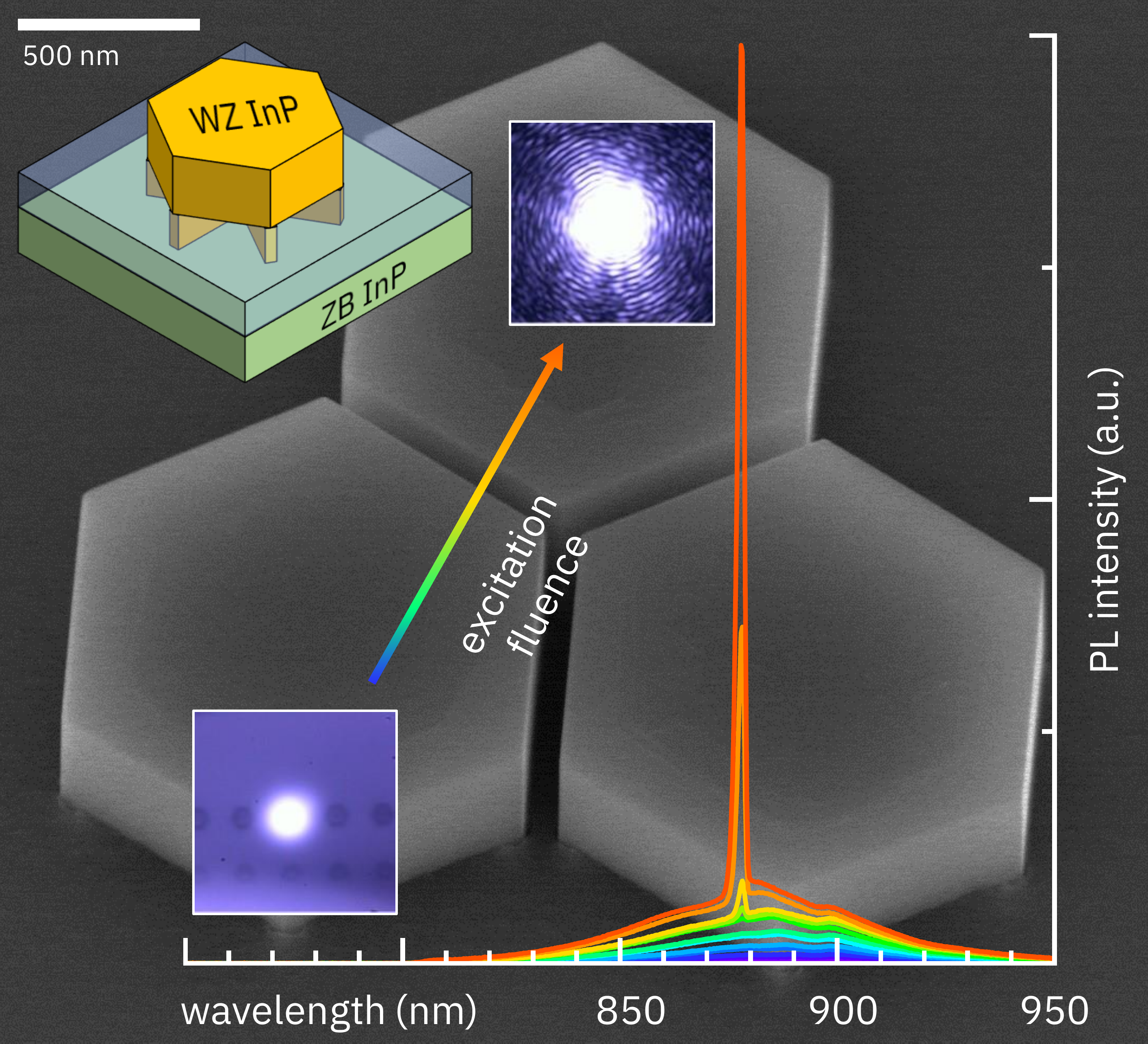}
	\vspace{-20pt}
\end{wrapfigure}

\textbf{Metastable wurtzite crystal phases of semiconductors comprise enormous potential for high-performance electro-optical devices, owed to their extended tunable direct band gap range. However, synthesizing these materials in good quality and beyond nanowire size constraints has remained elusive. In this work, the epitaxy of wurtzite InP microdisks and related geometries on insulator for optical applications is explored. This is achieved by an elaborate combination of selective area growth of fins and a zipper-induced epitaxial lateral overgrowth, which enables co-integration of diversely shaped crystals at precise position. The grown material possesses high phase purity and excellent optical quality characterized by STEM and \textmu-PL. Optically pumped lasing at room temperature is achieved in microdisks with a lasing threshold of 365 \textmu J/cm\textsuperscript{2}, thus demonstrating promise for a wide range of photonic applications.}

\vspace{0.3cm}

\begin{multicols}{2}

\begin{figure*}[t]
	\centering
	\includegraphics[width=\textwidth]{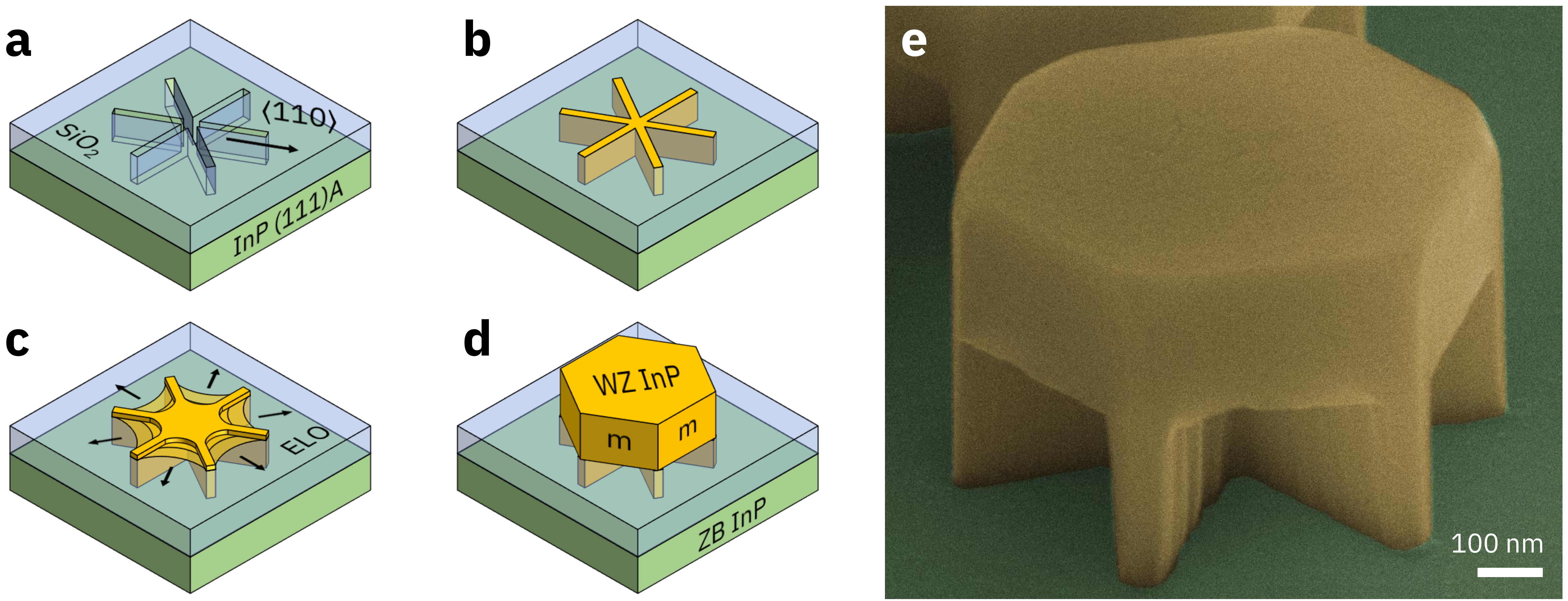}
	\caption{\textbf{Concept for WZ InP microdisk epitaxy.} (a) $\langle$110$\rangle$-oriented lines are patterned on a standard InP (111)A substrate covered with 300 nm SiO\textsubscript{2} by e-beam lithography and dry etching. (b) Growth is initiated in a fin-mode which allows for crystal phase switching to pure WZ. (c) Zipper-induced ELO maintains the metastable phase and provides for the formation of (d) hexagonally shaped microdisks with flat crystalline m-plane sidewalls. (e) False-colored 60° tilted SEM image of a typical structure with 900 nm diameter after removing the SiO\textsubscript{2} layer, revealing the pedestal fin structure. }
	\label{fig1}
\end{figure*}

\begin{figure*}[t]
	\centering
	\includegraphics[width=\textwidth]{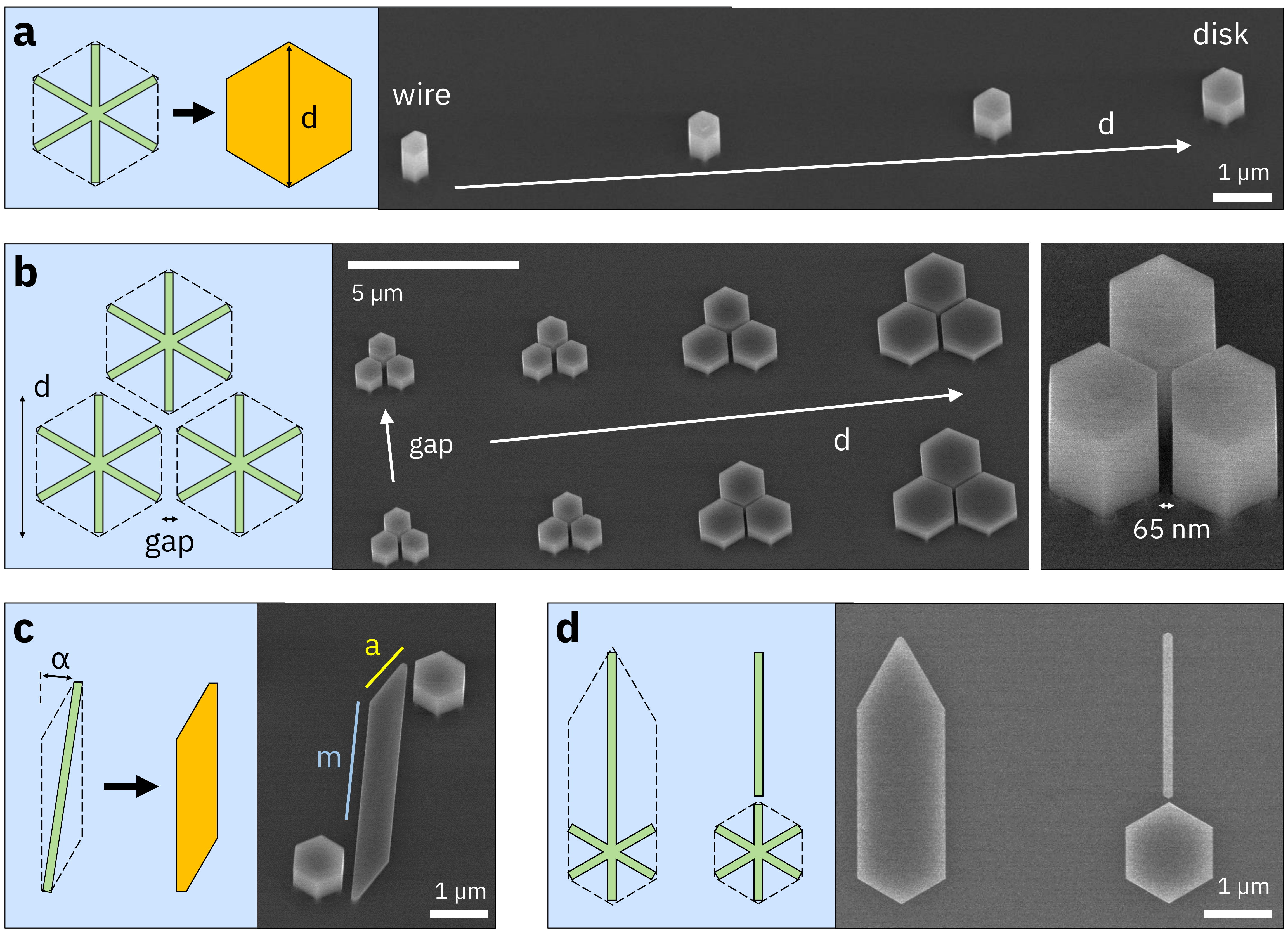}
	\caption{\textbf{Growth of structures with different feature sizes and geometries applicable to photonic devices.} (a) Changing the diameter of patterned stars allows to grow a range of feature sizes from wires to disks. (b) Multiple disks can be placed closely together to obtain arrays with tunable gap sizes. (c) Angular misaligned lines extend to an enclosing rhomboid formed from m- and a-planes. (d) The difference of a line opening connected and in proximity to a hexagon on the resulting crystal shape.}
	\label{fig2}
	\vspace{-1mm}
\end{figure*}

\begin{figure*}[t]
	\centering
	\includegraphics[width=\textwidth]{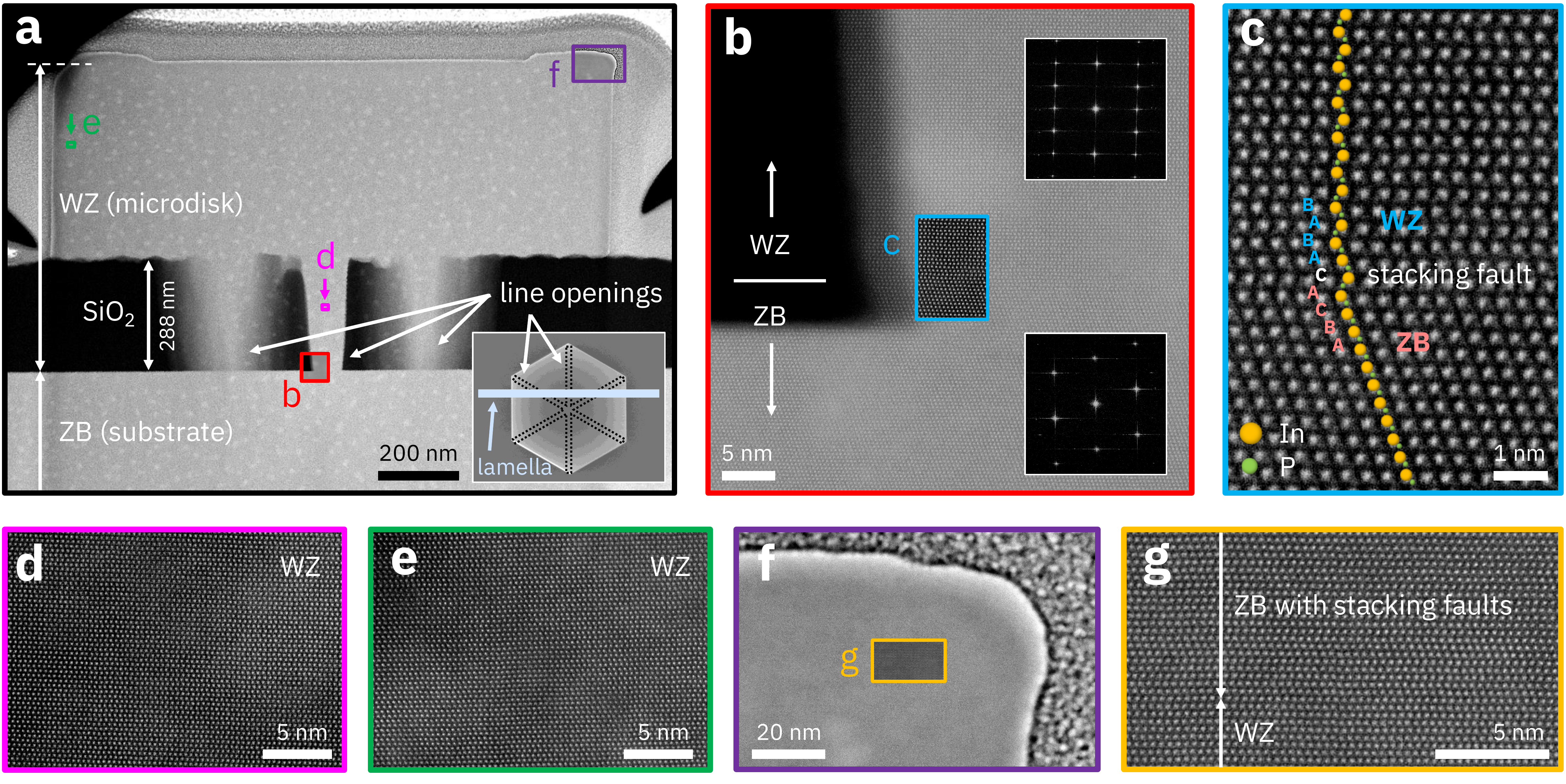}
	\caption{\textbf{HAADF-STEM characterization of a representative microdisk.} (a) FIB cross section of a microdisk with a diameter of 1.5 \textmu m and a height of 475 nm. The inset shows a top view image of the approximate position of the lamella with respect to the hexagon. Red, pink, green and purple marked areas indicate regions of interest, which are investigated in more detail. (b) High-resolution image at the central opening reveals a transition from ZB to WZ phase 5 nm from the substrate surface. Fast Fourier transform (FFT) images substantiate this observation. The area framed in blue is magnified in (c), which depicts the stacking change from ZB to WZ phase. A single stacking fault is obtained between otherwise phase-pure regions. (d)-(e) High-resolution images show pure WZ phase at various positions of the microdisk. (f-g) Magnification of the top right corner, showing faulted ZB material for the last ~30 nm of the growth.}
	\label{fig3}
\end{figure*}

\begin{figure*}[t]
	\centering
	\includegraphics[width=\textwidth]{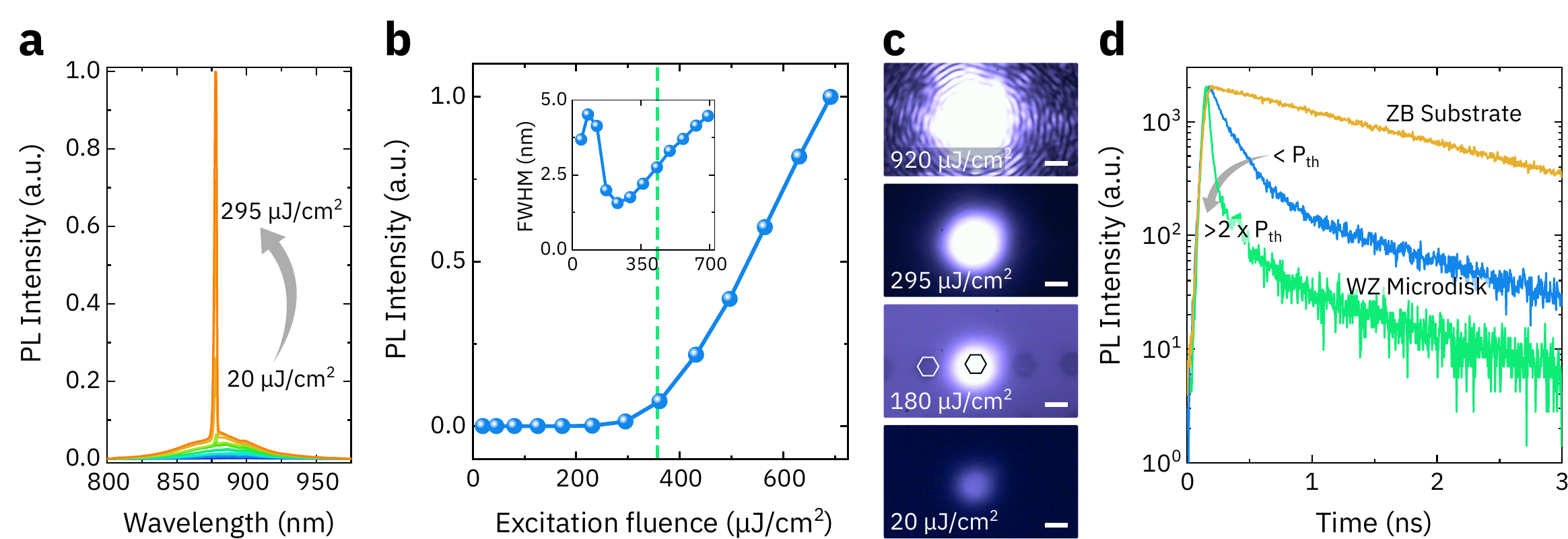}
	\caption{\textbf{Optical characterization.} (a) PL spectrum of a 2 \textmu m WZ microdisk for increasing optical excitation at the onset of lasing. (b) Linear LL curve of the device measured in (a). The inset shows the FWHM of the resonant mode versus excitation fluence. The dashed line depicts the laser threshold. (c) Far field images of the device in (a) at different excitation fluences. The image at 180 \textmu J/cm\textsuperscript{2} is taken under white light illumination. The hexagonal shapes indicate the fabricated devices. The scale bar in all far field images corresponds to 2 \textmu m. (d) TCSPC measurements of the ZB substrate, and a microdisk below and above threshold (>2$\times$P\textsubscript{th}).}
	\label{fig4}
\end{figure*}

III-V materials have attracted great interest throughout the past decades, due to their unique optical and electronic properties. The direct and tunable bandgap allows for scaled electro-optical devices with efficient emission and absorption properties. Novel concepts like on-chip optical communication~\cite{Miller:17}, complex optical networks~\cite{Gaio2019}, topological photonics~\cite{Zhao2018}, or metasurfaces~\cite{Liu2016} have been proposed and rely on the unique features of scaled III-V devices such as microdisk lasers~\cite{Wirths2018} or photonic crystals~\cite{Akabori2003}. Using a top down approach based on wafer bonding, lithography and etching, advanced III-V nanostructures on insulator can readily be formed for photonic applications~\cite{Crosnier2017,Morthier2015,DeGroote2016,Roelkens2007}. These approaches however, come at the cost of fabrication complexity and surface damage created during the etching process which negatively impacts device performance. An alternative approach is the direct monolithic growth of semiconductor micro- and nanostructures~\cite{Mayer2016,Wang2015,Kunert2016,Lourdudoss2012,Han2019}. Advancements in epitaxy techniques recently enabled the synthesis of high quality single crystalline material in various shapes, like nanowires~\cite{Tomioka2011}, nano- and microfins~\cite{Seidl2019}, microdisks~\cite{Mauthe2019}, or microrings~\cite{Mayer2019}. In contrast to semiconductor geometries obtained from etching, these structures are enclosed by as-grown crystal facets with superior quality. Moreover, materials that are not available in bulk form, like metastable wurtzite (WZ) III-Vs, can be grown and investigated~\cite{Gao2014b,Li2015,Kitauchi2010}. Synthesizing these materials in their thermodynamically less stable phase is challenging, but highly interesting in terms of optical properties. For instance, the entire composition spectrum of WZ InAlGaP shows a direct band gap transition, in contrast to their natural zinc-blende (ZB) counterparts~\cite{De2010,Assali2013a,Gagliano2018}.  Hence, metastable III-V materials are seen as a potential solution for the long-standing problem of realizing efficient green LEDs and laser diodes, commonly referred to as “green gap”~\cite{AufDerMaur2016a}.  Using conformal epitaxy or zipper-induced epitaxial lateral overgrowth (ELO) techniques, we demonstrated the growth of extended and pure WZ InP layers with layer dimensions ranging from <1 \textmu m\textsuperscript{2} to >100 \textmu m\textsuperscript{2}~\cite{Staudinger2018a,Staudinger2020}. In addition, the growth of various WZ microstructures, such as membrane-, prism-, and ring-like shapes by selective area epitaxy was recently shown~\cite{Wang2019}.

In this work, we present a versatile optical platform by demonstrating the growth of various InP micro- and nanostructures on an InP substrate with 300 nm oxide to provide optical isolation from the substrate. The structures are grown in a single growth run and thorough optical characterization is performed. The platform allows for exact positioning of epitaxially grown hexagons and related micro- and nano-sized shapes enclosed by a subset of the low-energy m- and a-plane facets, such as triangles, rhombi, fins, stars, or wires. Complex arrangements, e.g. arrays of hexagons with diameters ranging from 500 nm to 5~\textmu m and precise spacing can be achieved. Since the epitaxially grown crystals are only connected to the substrate via thin fins and otherwise separated by an oxide layer, optical isolation is achieved. 

Devices are fabricated using a three-step growth sequence: (1) inducing the transition of crystal phase to WZ, (2) anisotropic growth of WZ-fins and (3) zipper-induced ELO. The main steps of the formation process for a cavity of a microdisk laser are sketched in Figure~\ref{fig1}(a)-(d). A standard InP(111)A substrate is covered by 300 nm PECVD SiO\textsubscript{2} after which narrow (\textasciitilde 50 nm) lines are patterned by e-beam lithography and dry-etching techniques (see \nameref{Experimental} for details on the fabrication). The diameter of the microdisk will be determined by the length of the lines which are connected centrally and follow the three equivalent $\langle$110$\rangle$-directions. Metal-organic vapor phase epitaxy (MOVPE) allows for the selective nucleation of InP in the trenches and switching to WZ phase, followed by a highly selective and vertical growth along the [111]A direction. After the fins extend out of the oxide layer, the zipper-points at the center induce an ELO process until the stable \{1-100\} (m-plane) or metastable \{11-20\} (a-plane) WZ facets are formed~\cite{Wang2019}. Figure~\ref{fig1}(e) depicts a scanning electron microscope (SEM) image of a typical InP microdisk obtained during a growth time of 5 min, and after removing the oxide masking layer to reveal the underlying pedestal fin structure. The resulting microdisk has noticeable lower sidewall surface roughness compared to the underlying fin structures as well as to InP structures obtained by reactive ion etching processes.

The growth technique offers interesting options for designing a specific photonic structure, but also combinations of structures that can be grown in a single step. This includes hexagons of different sizes which can resemble either (nano-)wires or (micro-)disks. Crystal heights are mainly determined by deposition time, but also by diffusion mechanisms, which results in higher growth rates for hexagons with smaller diameter as illustrated in Figure~\ref{fig2}(a). Moreover, the resulting features are accurately defined by lithography and the respective resolution limit, including gap sizes as demonstrated in Figure~\ref{fig2}(b). Periodic arrays consisting of multiple structures, like hexagons, can be formed. Additionally, a wide range of different geometries and polygons can be grown together with no additional fabrication steps after epitaxy. This principle is exemplified in Figure~\ref{fig2}(c) and (d). Angled lines expand to rhomboids~\cite{Wang2019} that are enclosed by low energy facets and form structures that could serve as waveguides. Similarly, an elongated hexagon can be obtained by extending one of the seed lines. In principle, any convex shape can be envisioned which can be formed by m- and a-type facets, including for example triangles and rectangles. The well controlled and deliberate expansion of the exemplified structures beyond the width of the underlying fin together with the thick oxide mask layer are key for optical isolation and photonic functionality.

To obtain high quality optical devices from InP nanostructures, excellent crystalline quality needs to be achieved. Hence, we investigate the crystal structure and phase purity of the grown structures. Figure~\ref{fig3}(a) depicts a cross-sectional scanning transmission electron microscope (STEM) image of a typical microdisk with 1.5 \textmu m diameter imaged along a $\langle$110$\rangle$ zone-axis. Substrate and microdisk are divided by an oxide layer (dark), which is interrupted by three vertical bright stripes. These correspond to the line openings (compare Figure~\ref{fig1}(a)), as they are observed in a lamella which is slightly off-centered with respect to the hexagon-axis (inset of Figure~\ref{fig3}(a)). In order to investigate the crystal phase and defects we perform HR-STEM characterization in Figure~\ref{fig3}(b)-(g). We observe a transition from ZB to WZ phase after the crystal has extended approximately 5 nm into the oxide template. This is attributed to the initial growth stage at lower temperature, combined with the surface roughness of the substrate. At closer inspection the abruptness of this phase transition is revealed. The high-angle annular dark field (HAADF) image in Figure~\ref{fig3}(c) suggests that only a single bilayer (stacking fault) divides a pure ZB (ABCABC…) from a pure WZ (ABAB…) stacking sequence. Following this transition, fault-free WZ phase is observed (exemplified in two additional HR-STEM images) up to the flat top facet in the central area. Close to the side-facets, a ring-like structure terminates the top-surface, which is approximately 30 nm high. In stark contrast to the phase purity of the microdisk below, this ring shows ZB phase with high amount of stacking faults as indicated in Figure~\ref{fig3}(g). We assign this to the continued growth at reduced temperatures during the first seconds of the cool-down (see~\nameref{Experimental}).

In order to assess optical properties, the devices are characterized using a \textmu-photoluminescence (PL) setup with a ps-pulsed supercontinuum laser at 750 nm wavelength. A 100× objective is used to illuminate the devices from the top and collect their optical response. Figure~\ref{fig4}(a) depicts the photoluminescence response of a 2 \textmu m WZ microdisk for increasing excitation fluences (20 – 295 \textmu J/cm\textsuperscript{2}). At low illumination, the spontaneous emission of the microdisk is visible with a peak position of 880 nm corresponding to the bandgap energy of WZ InP and hence, confirming the crystal phase~\cite{Staudinger2018a}. Upon increasing laser power, a resonant mode starts forming at 875 nm. This can be observed in Figure~\ref{fig4}(b), where the integrated linear light in light out (LL) curve of the measured cavity mode is plotted. A clear kink marks the onset of lasing. By performing a linear fit on the right-hand side of the graph, the threshold of the resonant cavity mode can be determined to 365 \textmu J/cm\textsuperscript{2}. In logarithmic representation, a clear S-shape is revealed which is characteristic for a laser. Alongside with the strong increase of the emission in the resonant cavity mode, we observe a linewidth narrowing, another indication for lasing (see inset in Figure~\ref{fig4}(b)). The increase of the linewidth at high pump powers is due to a blueshift of the resonant wavelength arising from refractive index changes and is widely observed in similar systems~\cite{Bennett1990,Wang2015}. Additional insight can be obtained by analyzing far field images of the optical response taken with a standard camera. As shown in Figure~\ref{fig4}(c) the spontaneous emission is visible at low excitation fluences. With increasing pump energy, the emission of the devices gets stronger and fringes start forming. At high excitation fluences, a clear far-field radiation pattern with interference fringes is visible because of the extended first-order coherence of the laser emission. Room-temperature lasing operation in as-grown microdisks confirms the sufficient optical isolation provided by the 300 nm thick oxide layer introduced during fabrication, which was chosen for this first proof-of-concept.

To further quantify the material quality and the lasing performance, we perform time-correlated single photon counting (TCSPC) measurements. The sample is again illuminated with a 750 nm ps-pulsed supercontinuum laser using a 100× objective. Figure~\ref{fig4}(d) depicts the results of the TCSPC measurements. At room temperature and low excitation fluence, the ZB substrate and WZ microdisk exhibit carrier lifetimes of 1.57 ns and 239 ps, respectively. This is in agreement with results obtained in a previous work~\cite{Staudinger2018a}, and is attributed to the higher oscillator strength of the optical transition at the $\Gamma$-point in WZ phase~\cite{PhysRevB.73.235308,Wilhelm2012}. Under strong excitation power, the measured carrier lifetime of the WZ microdisk reduces strongly resulting in a value below the resolution limit of the measurement setup (<50 ps). This strong reduction of carrier lifetime is attributed to the stimulated emission in the resonant mode observed under high excitation powers, which further confirms lasing operation of the hexagonal microdisk with laser pulse durations below <50 ps. This constitutes, to the best of our knowledge, the first demonstration of a WZ InP microdisk laser. In combination with the possibility to controllably grow a range of different geometries side-by-side, our platform could have many promising future photonic applications.

In conclusion, we developed a methodology based on selective area epitaxy specifically tailored for the fabrication of photonic devices. High resolution lithographic patterning and thick oxide masks enable precise positioning and optical isolation, while zipper-ELO leads to large devices with defect-free and smooth crystal surfaces. Besides hexagonal microdisks we show a range of additional shapes to exemplify the capabilities of our platform. The unique fabrication approach allowed for the demonstration of optically driven lasing from hexagonal WZ InP cavities with a threshold of 365 \textmu J/cm\textsuperscript{2}. Finally, the concepts introduced are generic and could therefore also find applications in other commercially important material systems such as InAlGaP and III-nitrides.

\subsection*{Experimental Section}
\label{Experimental}

\textbf{Substrate fabrication:} An InP(111)A wafer was covered with nominally 300 nm plasma-enhanced chemical vapor deposited (PECVD) SiO\textsubscript{2} (288 nm measured during STEM). This thickness provides a compromise between sufficient optical isolation of the optical mode in the III-V material and fabrication simplicity. Line openings with widths of approximately 50-100 nm and varying lengths were patterned along the three equivalent $\langle$110$\rangle$-directions by e-beam lithography (EBL) and reactive ion etching (RIE) in a CHF\textsubscript{3}/O\textsubscript{2} plasma. Prior to growth the substrate was cleaned in acetone, isopropanol and O\textsubscript{2} plasma.

\textbf{MOVPE growth:} InP growth was carried out in a cold-wall showerhead MOCVD reactor using H\textsubscript{2} carrier gas, tertiarybutylphosphine (TBP) and trimethylindium (TMIn) at a total pressure of 8000~Pa. Partial pressures for TBP and TMIn were 8.2 Pa and 82 mPa, respectively, resulting in a nominal V/III ratio of 100. Before growth, the reactor was heated to 630 °C in a TBP atmosphere. Subsequently, deposition was initiated by introducing TMIn into the chamber. The substrate temperature was ramped up to 640 °C within the first 60 s, the final growth temperature. This procedure was developed in order to prevent substrate desorption prior the nucleation process. Deposition time was between 5 min and 30 min. Rapid cooling was started for the last 50 s of the growth, after which TMIn supply was terminated and cooling proceeded in TBP atmosphere.

\textbf{Optical characterization:} Optical characterization is performed using a \textmu-photoluminescence (PL) setup with a ps-pulsed supercontinuum laser at 750 nm (78 MHz repetition rate). The sample is illuminated from the top using a 100× objective (NA 0.6) which also collects the optical response of the sample. All measurements are performed at 300 K and under ambient conditions. The collected optical response is analyzed using a monochromator and an InGaAs CCD camera. Time-correlated single photon counting (TCSPC) measurements are performed in the same setup, using the 750 nm ps-pulsed supercontinuum laser and the 100× objective. A Si single photon detector is used to measure the optical response. Since the signal of the sample is very strong, a high-density filter was used to reduce the power of optical light incident on the detector. Using a TCSPC measurement module (PicoHarp 300), the lifetime of the sample can be determined. The resolution of the setup is ~50 ps due to the temporal pulse shape of the supercontinuum laser.

\subsection*{Acknowledgements}

The authors gratefully acknowledge Preksha Tiwari, Marilyne Sousa, Anna Fontcuberta i Morral and Heike Riel for fruitful technical discussions, as well as, the BRNC staff for technical support. The work presented here has received funding from the European Union H2020 program SiLAS (Grant Agreement No. 735008) and the ERC Starting Grant project PLASMIC (Grant Agreement No. 678567).

\subsection*{References}
\printbibliography[heading=none]

\end{multicols}
	
\end{document}